\documentclass[journal]{IEEEtran}

\usepackage{ifpdf}
\usepackage{cite}
\usepackage[pdftex]{graphicx}
\usepackage{subcaption}
\usepackage[dvipsnames]{xcolor}
\usepackage{marginnote}
\usepackage[textsize=small]{todonotes}
\usepackage{soul}
\usepackage{makecell}
\usepackage{ulem}
\usepackage{amsmath}
\begin{document}
%
\title{A Time and Cost Effective Method for Entropic Coefficient Determination of a Large Commercial Battery Cell}
%

\author{Zeyang Geng, Jens Groot,
        and Torbj{\"o}rn Thiringer,~\IEEEmembership{Senior Member, IEEE}
\thanks{Z. Geng and T. Thiringer are with the Department of Electrical Engineering, Division of Electric Power Engineering, Chalmers University of Technology, 412 96 Gothenburg, Sweden (e-mail: zeyang.geng@chalmers.se, torbjorn.thiringer@chalmers.se).}
\thanks{J. Groot is with Volvo Group Trucks Technology,
G{\"o}teborg, Sweden (e-mail: jens.groot@volvo.com).

\textbf{This paper has been accepted for publication by IEEE. DOI (identifier) 10.1109/TTE.2020.2971454. © Copyright 2020 IEEE. Personal use of this material is permitted. Permission from IEEE must be obtained for all other uses, in any current or future media, including reprinting/republishing this material for advertising or promotional purposes, creating new collective works, for resale or redistribution to servers or lists, or reuse of any copyrighted component of this work in other works.}}
}

\maketitle

\begin{abstract}

The entropic coefficient of a lithium-ion battery cell is used to calculate the reversible heat of a battery during operation, which is a non-negligible part for the battery thermal modelling. The contribution of this article is to propose a novel method to establish the entropic coefficient profile of a 26 Ah commercial pouch cell, and compare the results with those obtained from the traditional potentiometric and calorimetric methods, and all are found to be in a good agreement. The originality of this work is to use a method which consists of supplying a square pulse current wave form at a certain frequency and thus the resulting heat variation could be successfully linked to the input current using Fourier analysis. The current magnitude used were 1 C and 1.5 C, which are representative for the normal operation current in an electrified vehicle application. The method proposed is found to be cost efficient with a short experiment time and simple experiment setup. In fact, it can be used to characterize cells that are already mounted in a pack without the access to a climate chamber or calorimeter.
\end{abstract}


%
\IEEEpeerreviewmaketitle

\section{Introduction}
%
%
%
%
Electrification of vehicles as a response to the ambition of reducing tail-pipe emissions is today attracting a considerable interest. In order to alleviate the introduction of new electrified vehicles, a highly important issue is to prolong the life-time of the on-board batteries. One of the key factors for the battery life-time is the temperature control of the battery pack, as has been pointed out in several publications \cite{groot2015complex,broussely2005main,li2018temperature}. To control the battery pack temperature, a cooling and heating system is commonly used in the vehicles and an accurate thermal model needs to be developed to achieve a more efficient system design, where the heat source is one key feature. The heat source of a battery is not only the irreversible heat caused by the over potential but also the reversible heat caused by the entropy change \cite{al1999thermal,viswanathan2010effect,abdul2014thermal,skoog2017electro, 8433303, 6465668}. When the battery is charged slowly from a low state of charge (SoC) level, the temperature of the battery can even decrease due to the strong effect of the entropy change \cite{hong1998electrochemical}. Therefore, the entropy change needs to be taken into consideration to achieve a more accurate thermal model and in this way facilitate the design of the cooling system. Besides the improvement for thermal modelling, the entropy profile of a battery cell can also be used to study the battery aging \cite{7473682,zhang2018degradation} since it can reveal crystal structure changes in the electrodes \cite{maher2013effect,maher2014study,maher2014thermodynamic,osswald2015fast,yun2016analysis}.

The available methods to measure the entropy profile of a battery cell are the potentiometric method and calorimetric method. The potentiometric method was firstly introduced by Thompson in 1981 \cite{thompson1981thermodynamics} and has been widely used in the later research \cite{osswald2015fast,thomas2001measurement,zhang2017potentiometric}. In the potentiometric method the open circuit voltage of the battery is measured at different temperatures, which unfortunately is a very time consuming process. The open circuit potential follows a linear relationship with the battery temperature, and the slope of this linear relationship is the entropic coefficient $dU/dT$. Even with the voltage baseline correction \cite{osswald2015fast} it still requires 2.5 hours to acquire the value at a single SoC level for a commercial 18650 cell. The experiment time increases dramatically when applying the method on a larger commercial battery cell, not only because the thermal time constant increases due to a large thermal mass, but also an extra relaxation time is needed to obtain an accurate result.


Besides the potentiometric method, another option to measure the entropy change is the calorimetric method. The original purpose of the calorimeter used in the battery application is to study the total heat generation and there are two type of implementations: accelerating rate calorimeter \cite{hong1998electrochemical,al2000characterization,nieto2013thermal,vertiz2014thermal,schuster2015thermal,6563807,eddahech2013thermal} and isothermal heat conduction calorimeter \cite{xiao2013theoretical,chen2014measurements}. Since the calorimeter can only quantify the total heat generation, assistance from additional measurements is required to separate the reversible and irreversible heat. In the articles mentioned above, the reversible heat is estimated with the input from the potentiometric method whereas the irreversible heat is estimated from the battery impedance, which is obtained from electrochemical impedance spectroscopy measurement or current interruption measurement. There is a lack in the literature for using the calorimeter method directly to determine the entropic coefficient. The exception is Damay et al. \cite{damay2016method} who used the calorimeter as the direct measurement to determine the entropy profile with various current levels. In \cite{damay2016method} the entropy effect is calculated from the subtraction of the charging heat and discharging heat, therefore only the average entropy profile of charge and discharge can be obtained. Overall the calorimetric method is a relatively fast method and it gives a continuously entropy profile for the whole SoC range. However the considerable effort and high cost of the calorimeter equipment, especially for large battery cells, has become the main limitation of this method.


To sum up, the potentiometric method is time consuming and less applicable to characterize a large battery cell. Moreover, it also requires high precision potential measurement in a temperature controlled test environment. The calorimetric method is fast and accurate, but it requires an even more advanced experimental setup. As the state of the art review has shown, much valuable effort regarding the entropic coefficient determination has been done. However, what is missing and to pursue is a method that could be used after cells have been inserted in a pack, i.e. in-situ, would be of great value. 

The contribution of this article is to demonstrate the effectiveness of a novel method for determining the entropy effect with a minimum time and special equipment need. In this new method the determination can even be done in-situ in any battery pack as long as the battery pack is equipped with internal temperature sensors. Furthermore, another goal with the article is to compare the results of using the three methods. With this information it is possible to chose the most suitable method based on the application as well as on the available test equipment.

\section{Heat effect in a battery cell}

\begin{figure}[!ht]
    \centering
    \includegraphics{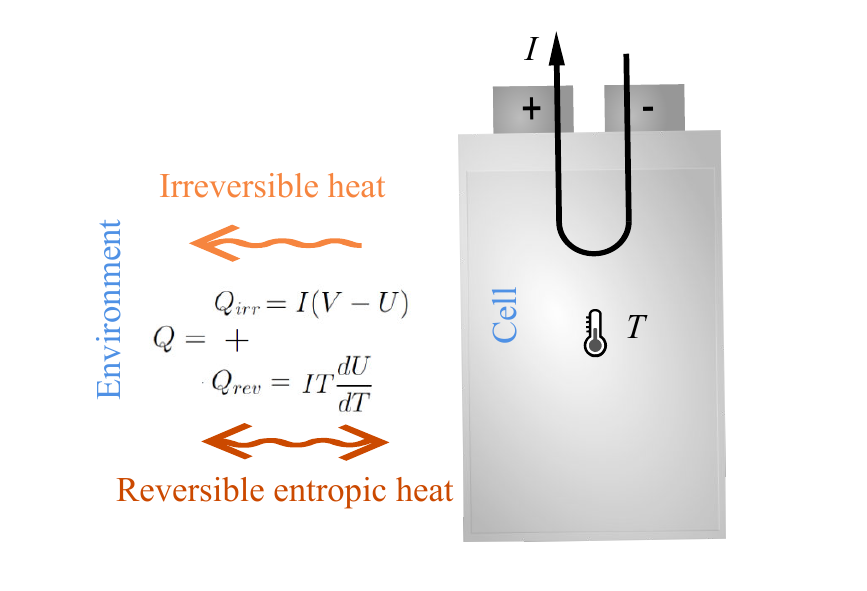} \\
    \caption{Heat generation in a lithium ion battery cell.}
    \label{fig:Equation_illustration}
\end{figure}

The total heat generation of a battery is made up by two parts, the irreversible heat $Q_{irr}$ caused by the over potential and the reversible heat $Q_{rev}$ caused by the entropy change \cite{pals1995thermal}, and their sum can be found as
\begin{equation}
    Q =  Q_{irr} + Q_{rev} = I(V-U)+IT\frac{dU}{dT}
    \label{eq:Heat_generation}
\end{equation}
where $I$ is the current passing through the battery, $T$ is the temperature of the battery, $V$ is the terminal voltage and $U$ is the open circuit potential of the battery cell in equilibrium state. In this article, the current $I$ is defined as positive when the battery is being charged and $Q_{rev}$ is positive when heat is dissipated from the battery system to the environment, as illustrated in Fig.~\ref{fig:Equation_illustration}.

In (\ref{eq:Heat_generation}) $dU/dT$ is the entropic coefficient which is related to the entropy change of the reaction by
\begin{equation}
    \frac{dU}{dT} = \frac{\Delta S}{nF} 
\end{equation}
where $\Delta S$ is the entropy change in the battery cell, $n$ is the number of electrons transferred per atom and $F$ is the Faraday constant \cite{newman2012electrochemical}. Depending on the sign of the current $I$ and the entropic coefficient $dU/dT$, the entropy heat can be either positive or negative. As a contrast, the overpotential of the battery leads to irreversible heat, i.e. losses from the battery. It is caused by the ohmic drop in the current collectors, contact resistance between the active particles, kinetics of charge transfer on the phase boundary and finally the mass transport of lithium ions in both electrolyte and solid particles. The overpotential depends non-linearly on the current magnitude, sign of the current, SoC and temperature etc. On the other hand, the reversible heat has a linear relation with the entropic coefficient $dU/dT$ which depends on SoC only. When the current is very low, the irreversible heat is not significant and the entropy effect dominates the total heat generation. With an increasing current, the irreversible heat plays a more and more important role and the entropy effect is less observable.

In this article we implemented three methods to determine the entropic coefficient $dU/dT$: potentiometric method, calorimetric method and the proposed novel method. The experiment setup and results are described in the following.

\section{Test object}

The battery used in this work is a 26 Ah commercial lithium-ion pouch cell. The positive electrode is a mixture of 70wt\% (weight percentage) LiNi$_{1/3}$Mn$_{1/3}$Co$_{1/3}$O$_2$ (NMC) and 30wt\% LiMn$_2$O$_4$ (LMO) and the negative electrode is graphite \cite{wikner2017lithium}. The battery cell profile is given in Table~\ref{tab:cellspecification}.


\begin{table}[!ht]
\centering
\caption{Cell specification.}
\begin{tabular}{| l| c | }  \hline
Capacity & 26 Ah \\ \hline
Max charge voltage & 4.15 V \\ \hline
Discharge cut-off voltage & 2.8 V \\ \hline
Internal impedance & 3 m$\Omega$\\
\hline
Dimension & 15 cm $\times$ 20 cm $\times$ 7 mm\\
\hline
\end{tabular}
\label{tab:cellspecification}
\end{table}

The entropy profiles of the individual positive electrode materials have been reported in \cite{shi2016effect} for NMC and \cite{thomas2001measurement} for LMO. Huang et al. also studies the entropy profile for a blended positive electrode with NMC and LMO mixed by a mass ratio 1:1 in \cite{huang2015entropy}. Graphite is one the most commonly used negative electrode materials for lithium ion batteries and its thermodynamic properties has been well-investigated by using both experimental methods \cite{reynier2003entropy, reynier2004thermodynamics} and theoretical analysis \cite{perassi2016theoretical}.

\section{Investigation of the proposed method to determine the entropic coefficient}

\begin{figure}[!ht]
     \centering
     \begin{subfigure}[b]{\columnwidth}
         \centering
         \includegraphics{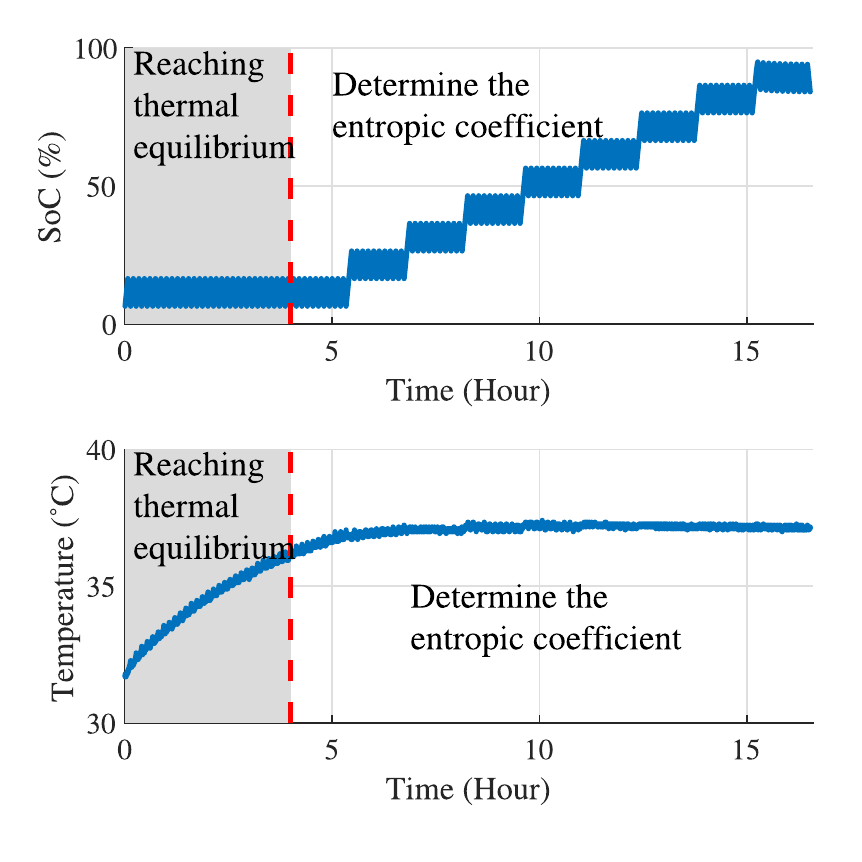}
         \caption{}
         \label{fig:Method3_measurement_thermal_eq}
     \end{subfigure}
     \begin{subfigure}[b]{\columnwidth}
         \centering
         \includegraphics{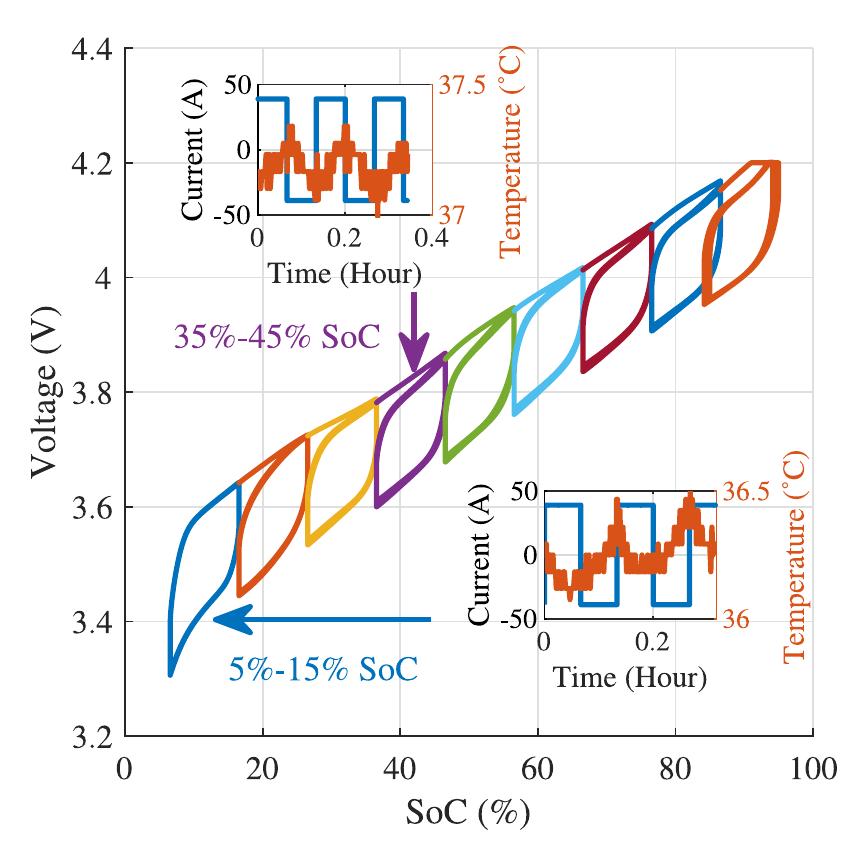}
         \caption{}
         \label{fig:Method3_measurement_4}
     \end{subfigure}
     \hfill
        \caption{Implementation of the alternating current method. (a) Before the actual test, the battery is cycled during the initial SoC interval to reach a thermal equilibrium state by utilizing the irreversible heat. (b) The temperature swings caused by the entropy change at different SoC intervals.}
        \label{fig:Method3}
\end{figure}

Both the commonly used potentiometric and the less commonly used calorimetric methods are known to provide good results of the entropic coefficient. However when it comes to the characterization of a large battery cell or when the cell is placed in a complete battery pack, the potentiometric and calorimetric methods are limited and less applicable due to practical reasons as mentioned in the introduction. To be able to measure the entropic coefficient of a large battery cell or in the case where a temperature chamber or calorimeter is not available, we here propose a new method by applying an alternating current to charge and discharge the battery within a small SoC interval, as shown in Fig.~\ref{fig:Method3}.

When the battery is being charged and discharged with a square  current wave $I$ within a small SoC interval, the irreversible heat $Q_{irr} = I^2R$ can be considered constant as the battery impedance $R$ can be considered constant. Meanwhile, the reversible heat $Q_{rev} = IT dU/dT$ is also a square wave with the same frequency as the input current. The constant irreversible heat keeps the system in the thermal equilibrium state (after t = 5 hours in Fig.~\ref{fig:Method3_measurement_thermal_eq}) and the alternating reversible heat leads to a temperature swing that reflects the entropy effect.

To be able to assume the battery impedance as a constant value during the SoC interval and avoid the temperature impact, the battery is kept in a thermal equilibrium state with the environment to keep a constant temperature. Therefore before the test, the battery has been cycled between the initial SoC interval 5 \% and 15 \% for 4 hours, as shown in Fig.~\ref{fig:Method3_measurement_thermal_eq}. During this initial cycling the irreversible heat is used to heat up the system to reach a thermal steady state and the following procedure is the actual entropic coefficient measurement.

The reversible heat generation can be treated as a linear system where the current is the input, $Q_{rev}$ is the output and $TdU/dT$ is the transfer function. With Fourier analysis the equation can be expressed as 
\begin{equation}
    \frac{1}{2}Q_{rev(0)} + \sum_{n=1}^{\infty} Q_{rev(n)} = T\frac{dU}{dT}( \frac{1}{2}I_{(0)} +  \sum_{n=1}^{\infty} I_{(n)})
\end{equation}
Under the condition of a constant battery temperature, if an input current signal with a certain frequency is applied to the system, the entropic coefficient can be identified in frequency domain as
\begin{equation}
     \frac{dU}{dT} = \frac{Q_{rev(1)}}{I_{(1)}T}
     \label{eq:system_identification}
\end{equation}
where $Q_{rev(1)}$ and $I_{(1)}$ are the fundamental components in the frequency domain of the output and input signal respectively. Ideally a sinusoidal current wave is the best input signal for system identification. However, in this work a square current wave is applied, as it can be generated more conveniently with standard equipment at a desired power level.

\begin{figure}[!ht]
    \centering
    \includegraphics{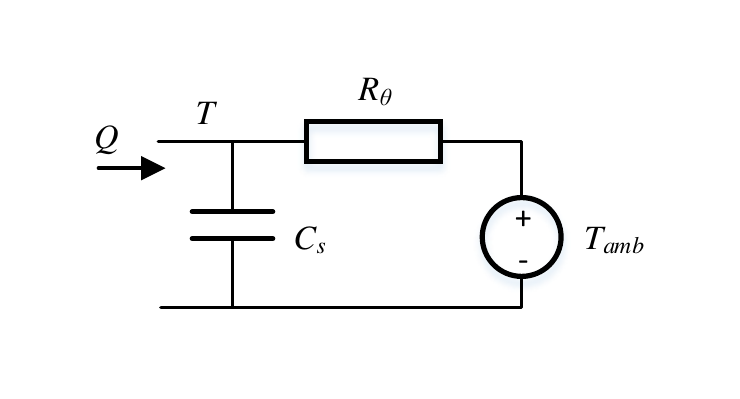} \\
    \caption{Thermal model used to estimate the entropic heat.}
    \label{fig:Equation_thermal_illustration}
\end{figure}

The heat $Q_{rev}$ generated during the experiment, which is the output signal in (\ref{eq:system_identification}), can be calculated from the temperature following the heat transfer equation
\begin{equation}
    Q =  C_s\frac{dT}{dt} + \frac{1}{R_\theta}(T-T_{amb})
    \label{eq:Heat_transfer}
\end{equation}
where $C_s$ is the heat capacity of the system, $R_\theta$ is the thermal resistance between the battery and environment, $T$ is the temperature of the battery and $T_{amb}$ is the ambient temperature as illustrated in Fig.~\ref{fig:Equation_thermal_illustration}.

If the thermal model of the system (including the battery and the surrounding fixture) reacts as a first order system, then the parameters $C_s$ and $R_\theta$ can be easily identified from a step response. To determine the thermal parameters $C_s$ and $R_\theta$ in (\ref{eq:Heat_transfer}), a calibration test is performed where a heating element is inserted next to the battery to apply a step power input with a known power level (1~W, 3~W and 5~W) until reaching almost steady state. The thermal capacity $C_s$ and thermal resistance $R_\theta$ are calculated to be 4225 J~K$^{-1}$ and 4.22 K~W$^{-1}$ respectively for the system used in this study. The most important parameter to identify in this method is the heat capacity $C_s$ since the alternating current period is far shorter than the thermal constant of the system (around 5 hours) and the thermal resistance $R_\theta$ affects the result very little. For a simplification, the heat capacity can also be estimated analytically as

\begin{equation}
\begin{split}
C_s =\sum_{i=1} C_{p,i}m_i & = C_{p,cell}m_{cell}+C_{p,fixture}m_{fixture} \\
  & ~~ + C_{p,Styrofoam}m_{Styrofoam}
\end{split}
\end{equation}
where $C_{p,i}$ and $m_i$ are the specific heat capacity and the weight of each component in the system, including the battery and fixtures.

\begin{figure}[!ht]
    \centering
    \includegraphics{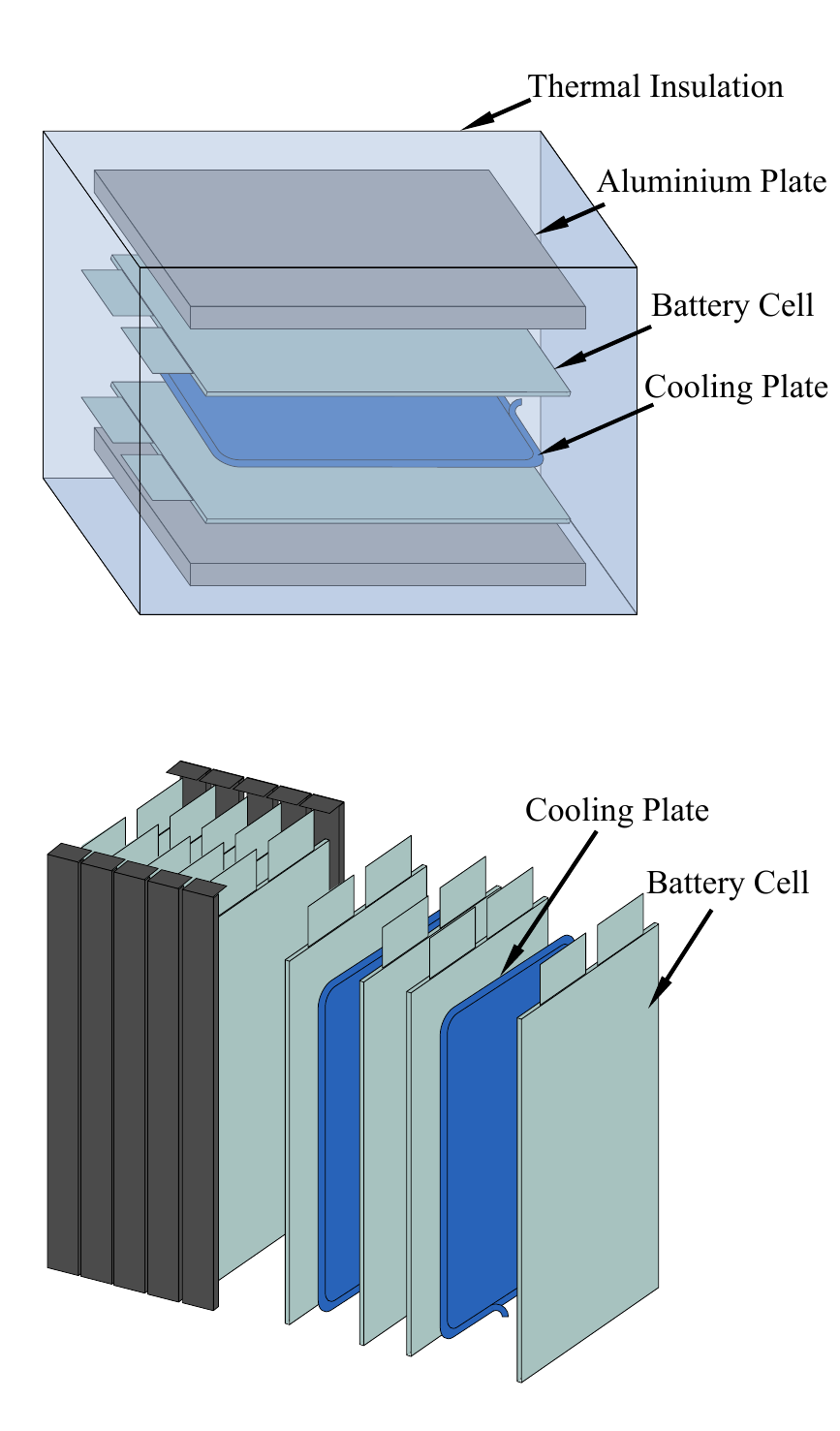} \\
    \caption{The experimental setup for the proposed method and one example of the battery cell assembly in a pack.}
        \label{fig:Method3Setup}
\end{figure}

In the experiment for this method, the cell tester is a MACCOR 4000 with 0.02\% of full scale voltage and 16 bits resolution. The temperature is recorded with type K thermocouples and the sampling frequency is 1 Hz. During the experiment, the cells are located between two aluminium plates with certain pressure applied and the test jig is thermally insulated. This is to emulate the in-situ environment where the battery cells are assembled in a pack. There are various configurations of the battery packs and one possible example is shown in Fig.~\ref{fig:Method3Setup}.

\begin{figure}
    \centering
    \includegraphics{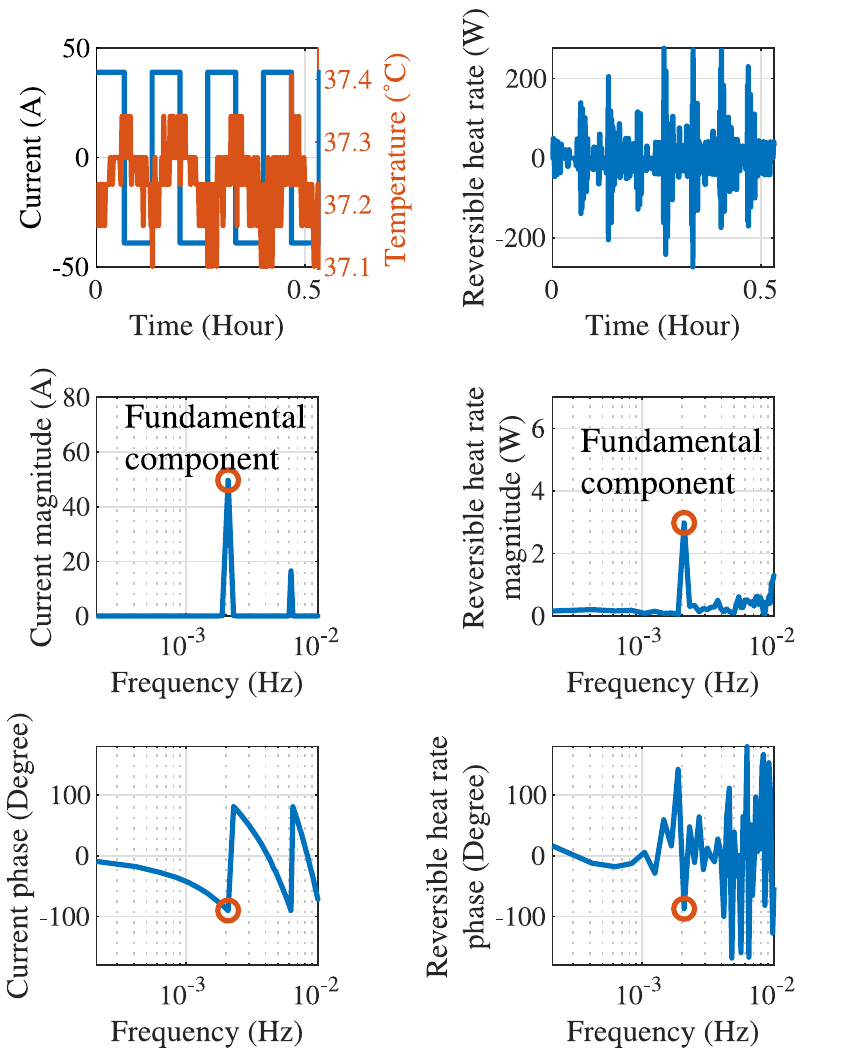} \\
    \caption{An example of the input current and its resulting reversible heat rate in time and frequency domain.}
    \label{fig:Method3_result_fft}
\end{figure}

With the proposed method, the entropic coefficient is measured for 9 SoC levels from 5\% SoC to 95\% SoC with 10\% SoC intervals. During the measurement, the battery starts at 5\% SoC and a square current wave with a 1 C amplitude and 12 minutes cycling period are applied for 10 cycles at each SoC interval. After the cycling at one SoC level, an additional charge pulse is applied to go to the next SoC level. When the battery reaches 95\% SoC, the same procedure is repeated but with an additional discharge pulse to change the SoC level. As it is shown in Fig.~\ref{fig:Method3_measurement_4}, at 5\%-15\% SoC interval, the negative entropic coefficient causes a temperature swing where the positive current corresponds to a decreasing temperature. The sub-figure shows the temperature behavior versus current within the small SoC interval. An opposite behaviour at 35\%-45\% can be observed due to the positive entropic coefficient.

A higher current wave of 1.5 C is also used with 8 minutes cycling period to check the repeatability. Since the entropic coefficient is frequency independent, the excitation current can have various cycling periods. A shorter cycling period can determine the entropic coefficient with a finer SoC step, however the temperature may not have enough time to build up and the result may be hidden in the noise.

\begin{figure}[!ht]
    \centering
    \includegraphics{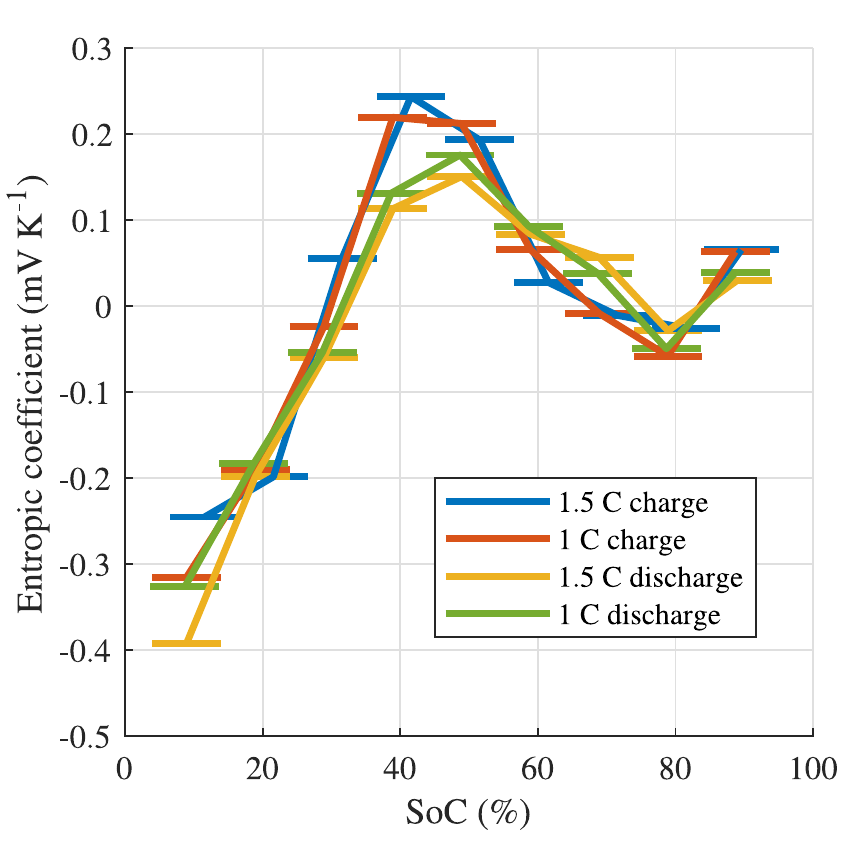} \\
    \caption{Entropic coefficient measured with the proposed method with an alternating current. The measurement are performed with 1 C and 1.5 C giving almost identical result.}
    \label{fig:Method3_result}
\end{figure}

Fig.~\ref{fig:Method3_result_fft} shows an example of how Fourier analysis is applied to determine the entropic coefficient. Although the noisy temperature measurement makes the estimated reversible heat difficult to identify in the time domain, the fundamental component in the signal stands out clearly in the frequency domain despite the high frequency noise, which makes the proposed method capable to measure the entropic coefficient with simple equipment. In this example the current wave has a cycling period of 8 minutes which corresponds to a 2.08 mHz fundamental frequency ($f=1/T$). The signal magnitude and phase at the fundamental frequency represents the fundamental component in (\ref{eq:system_identification}). In the selected SoC interval the fundamental components of the current and the reversible heat are in phase, which indicates a positive entropic coefficient, i.e. the temperature will increase with a positive current. Vice versa, if the entropic coefficient is negative, the reversible heat will have a 180 degree phase different from the current.

The result from this method is shown in Fig.~\ref{fig:Method3_result}. The measurements with different C-rates give almost identical results except for minor spreading at the very low SoC range. However the pattern of the entropic coefficient shifts slightly between the charge and discharge, exactly as it is the case of the calorimetric method measurement in Fig.~\ref{fig:Method2_result}.

\begin{figure}[!ht]
    \centering
    \includegraphics{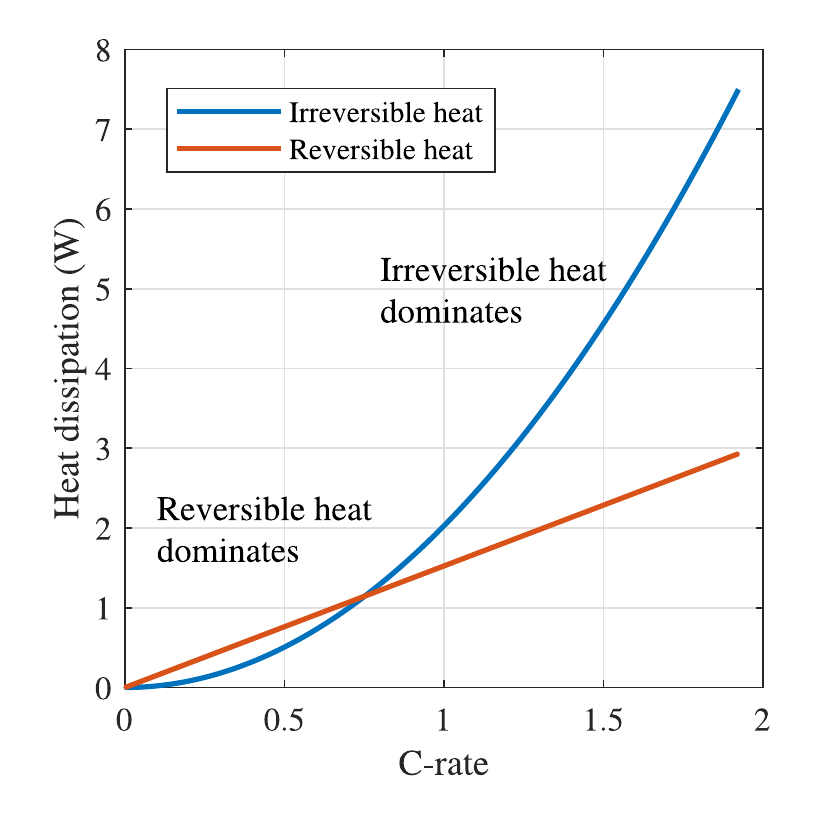} \\
    \caption{Relation between reversible heat and irreversible heat depending on the current level.}
    \label{fig:heat_current}
    
\end{figure}

Not only is the proposed method accurate and cost efficient, but also it studies the battery entropy effect under the condition when it is placed in operation, for instance in a vehicle application. In the potentiometric method, there is no net current flowing through the battery during the measurement and in the calorimetric method, the currents used for test is 0.1 C and 0.05 C. However, when the battery is used in the vehicle application, the root-mean-square (RMS) current is normally higher than 1C for a plug-in hybrid electric vehicle (PHEV). Therefore, there is a substantial interest to investigate whether the entropic coefficient identified with smaller currents is valid in the operation condition with higher currents. As discussed previously, with the current increasing, the irreversible heat increases and dominates the total heat dissipation, as shown in Fig.~\ref{fig:heat_current}. In this example, the battery internal resistance is 3 m$\Omega$ according to a previous study \cite{Chaudhry2018investigate} and the entropic coefficient is 0.2 mV~K$^{-1}$, corresponding to 50 \% SoC of the battery being tested. The calculation is made for a room temperature 20~$^{\circ}$C. At a lower current level, the reversible heat can be larger than the irreversible heat so it is important to have an accurate entropic coefficient profile for the thermal modelling. However at a higher current level the irreversible heat is dominating and thus the reversible heat effect might be difficult to observe directly.

\section{Determination of the entropic coefficient using traditional methods}

\begin{figure*}[!h]
     \begin{subfigure}[b]{\columnwidth}  
         \centering
         \includegraphics{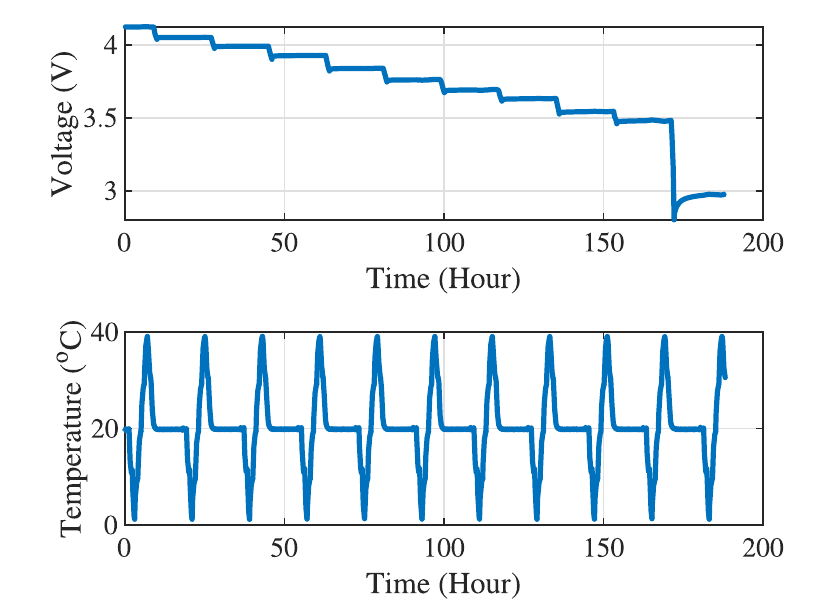}
         \caption{}
         \label{fig:Method1_setup}
     \end{subfigure}
     \hfill
     \begin{subfigure}[b]{\columnwidth}  
         \centering
         \includegraphics{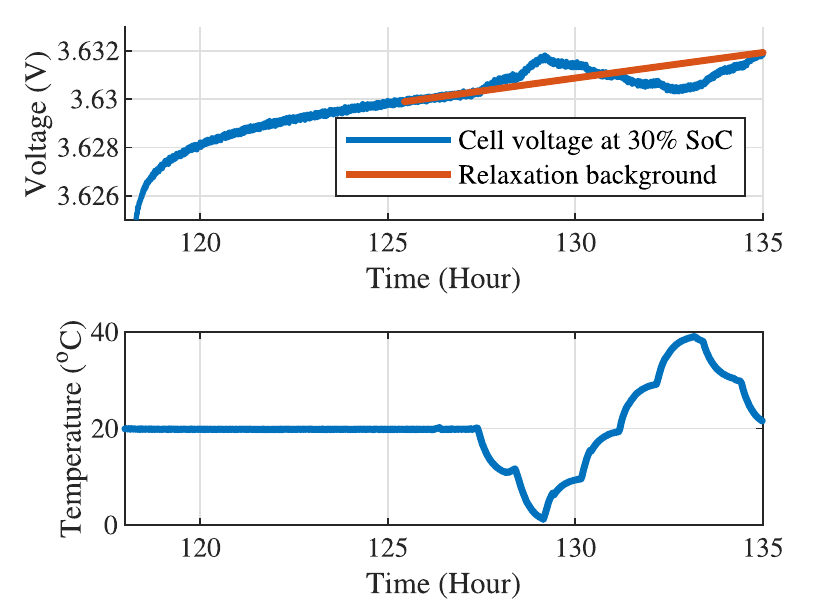}
         \caption{}
         \label{fig:Method1_zoomin}
     \end{subfigure}
     \hfill
     \newline
     \begin{subfigure}[b]{\columnwidth}  
         \centering
         \includegraphics{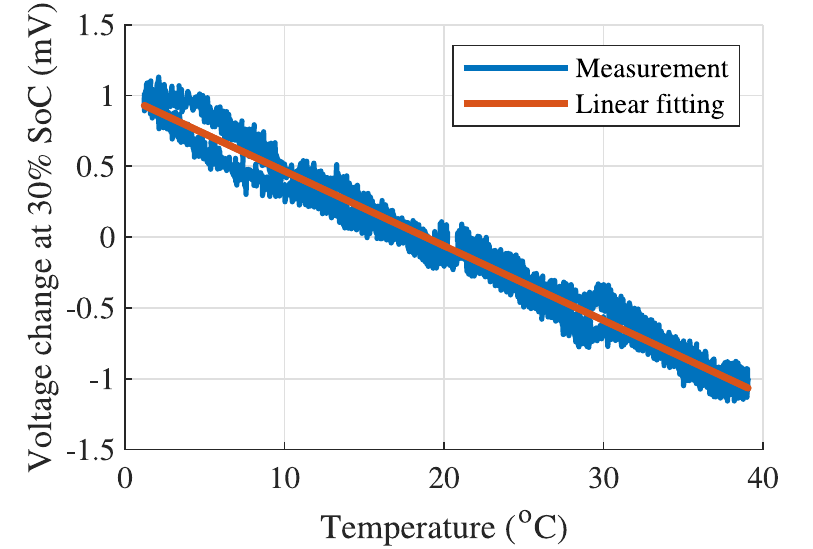}
         \caption{}
         \label{fig:Method1_fitting}
     \end{subfigure}
     \hfill
     \begin{subfigure}[b]{\columnwidth}  
         \centering
         \includegraphics{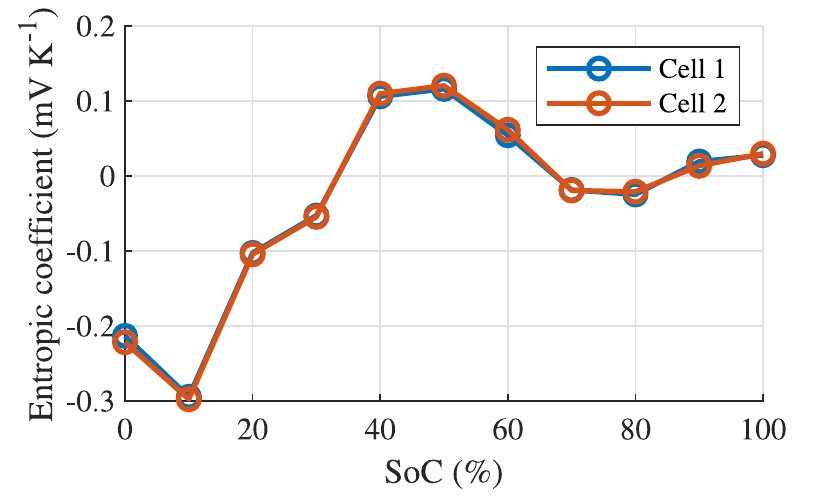}
         \caption{}
         \label{fig:Method1_result}
     \end{subfigure}
     \hfill
     
        \caption{Potentiometric method implementation and its result. (a) The complete test procedure. (b) An example of the voltage response at different temperatures. The discharge ended at t = 118 hour. The relaxation background is estimated and subtracted to achieve an accurate result. (c) Linear fitting to obtain the entropic coefficient. (d) Entropic coefficient result from the potentiometric method at 11 SoC levels between 0 $^{\circ}$C and 40 $^{\circ}$C.}
        \label{fig:Method1}
\end{figure*}

In order to validate the results, the most commonly used method, the potentiometric method is used. In addition also the less commonly used, but still known methodology, the calorimetric method, is also used for the validation of the proposed novel method.

\subsection{Result when using the Potentiometric method}

As mentioned in the introduction, the open circuit potential of a battery cell in equilibrium state follows a linear relationship with the battery temperature and the slope of this linear relationship is the entropic coefficient $dU/dT$. To achieve an accurate result with the potentiometric method, a high precision voltage measurement is required and the cell needs to be at equilibrium state for each measurement point.

The experiment equipment used in the potentiometric test is a cell tester PEC ACT0550 with a programmable climate chamber Espec LU-124. The voltage measurement accuracy is $\pm$0.005\% Full Scale Deviation and the temperature is measured by a type K thermocouple with 0.01$^oC$ resolution. In this measurement it is the temperature change $\Delta T$ that affects the result but not the absolute temperature.

In the experiment, two battery cells are tested and they are located in a fixture with certain pressure applied. The entropic coefficient is measured at 11 SoC levels from 0\% to 100\% SoC with 10\% intervals. To start with, the battery cells are fully charged to 100\% SoC and then discharged to the next SoC level with 0.1C for one hour. After each discharge pulse, the cell is relaxed for 8 hours at 20 $^{\circ}$C, and then the entropic coefficient is measured by changing the climate chamber temperature between 0 $^{\circ}$C and 40 $^{\circ}$C with 10 $^{\circ}$C steps, as shown in Fig~\ref{fig:Method1_setup}. Due to the large thermal mass of the battery, the temperature of the climate chamber at each step is kept constant for one hour to reach a thermal equilibrium state. As can be noted in Fig.~\ref{fig:Method1_zoomin}, the battery voltage is still changing after 8 hours relaxation and a relaxation background subtraction is needed when calculating the entropic coefficient. The relaxation background is fitted in MATLAB with a polynomial of degree 2 as shown in Fig.~\ref{fig:Method1_fitting}.

\begin{figure*}[!h]
     \begin{subfigure}[b]{\columnwidth}  
         \centering
         \includegraphics[width=\columnwidth]{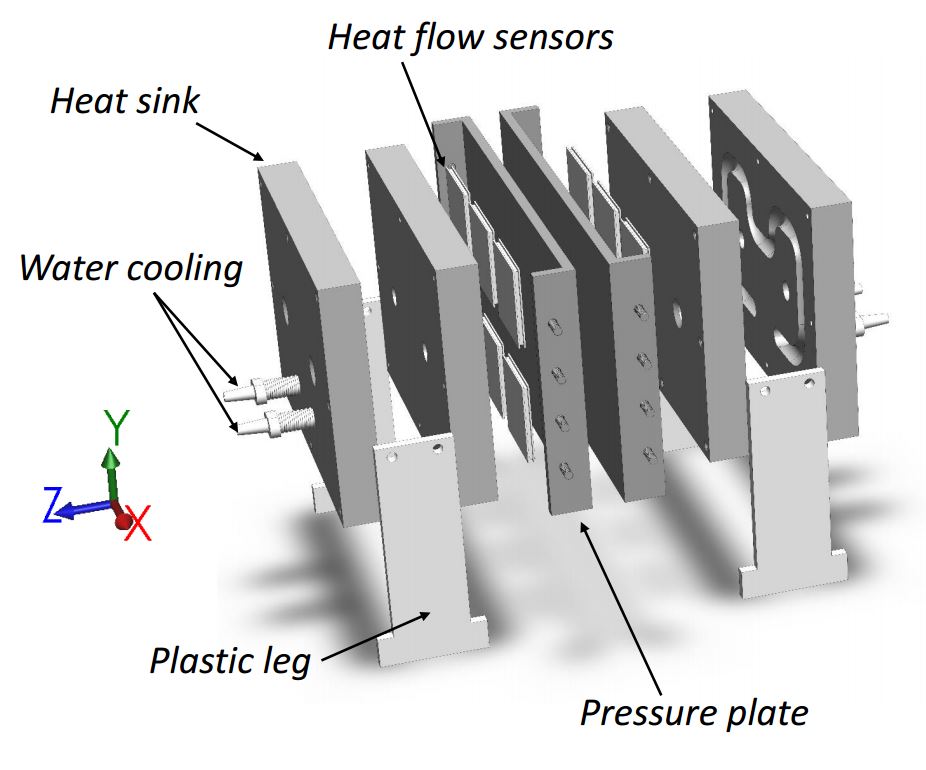}
         \caption{}
         \label{fig:calorimeter1}
     \end{subfigure}
     \hfill
     \begin{subfigure}[b]{\columnwidth}  
         \centering
         \includegraphics[width=\columnwidth]{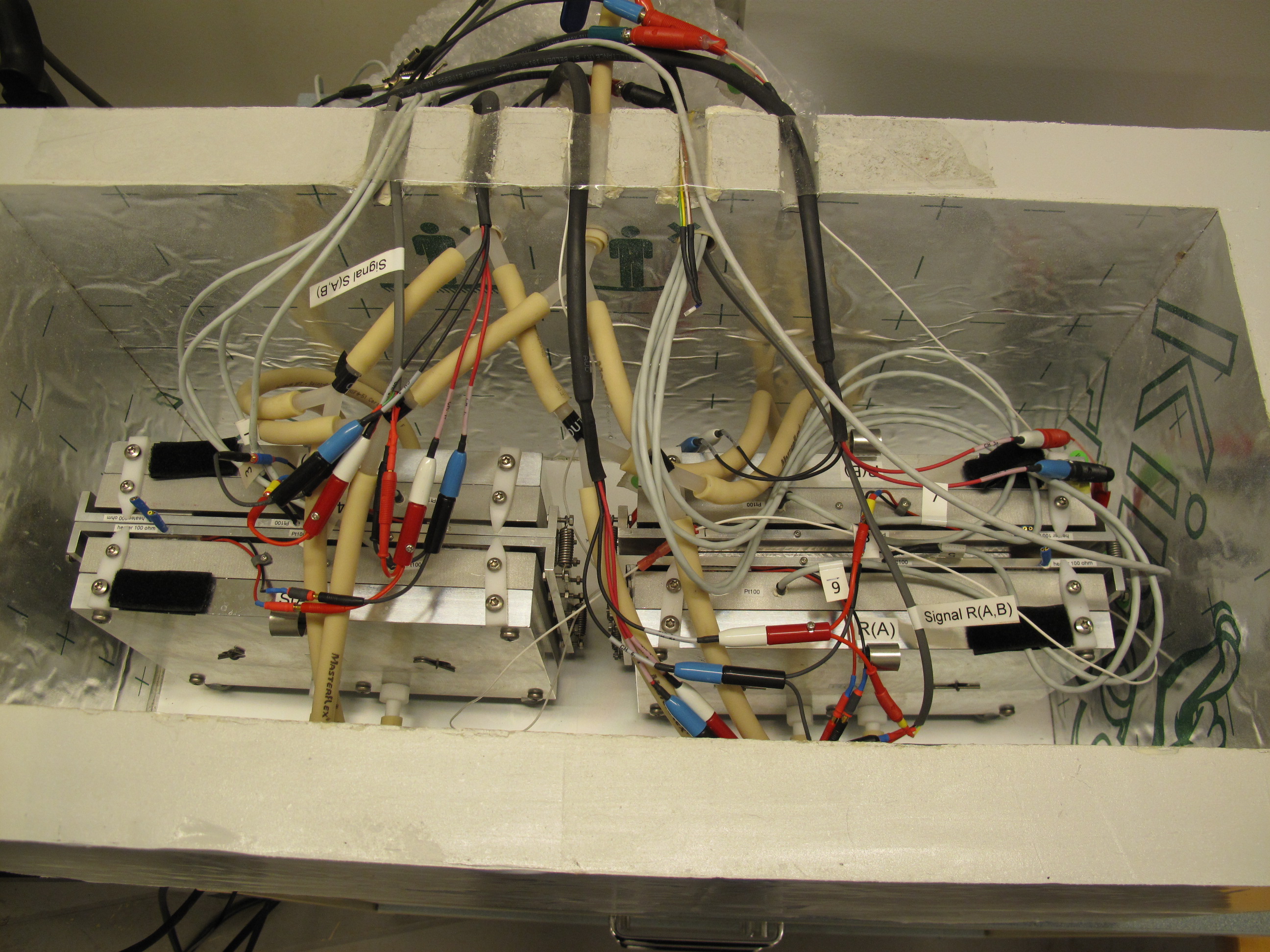}
         \caption{}
         \label{fig:calorimeter2}
     \end{subfigure}
     \hfill
     \newline
     \begin{subfigure}[b]{\columnwidth}  
         \centering
         \includegraphics{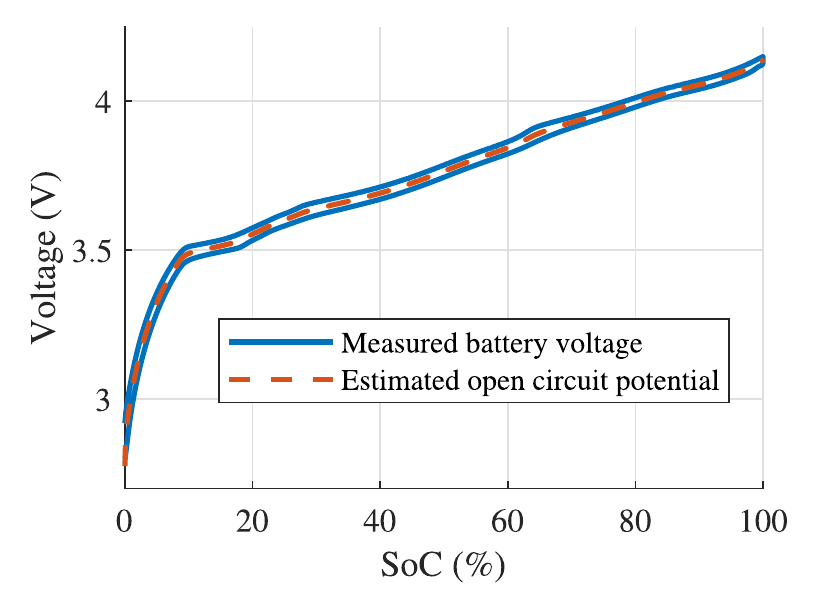}
         \caption{}
         \label{fig:Method2_ocv}
     \end{subfigure}
     \hfill
     \begin{subfigure}[b]{\columnwidth}  
         \centering
         \includegraphics{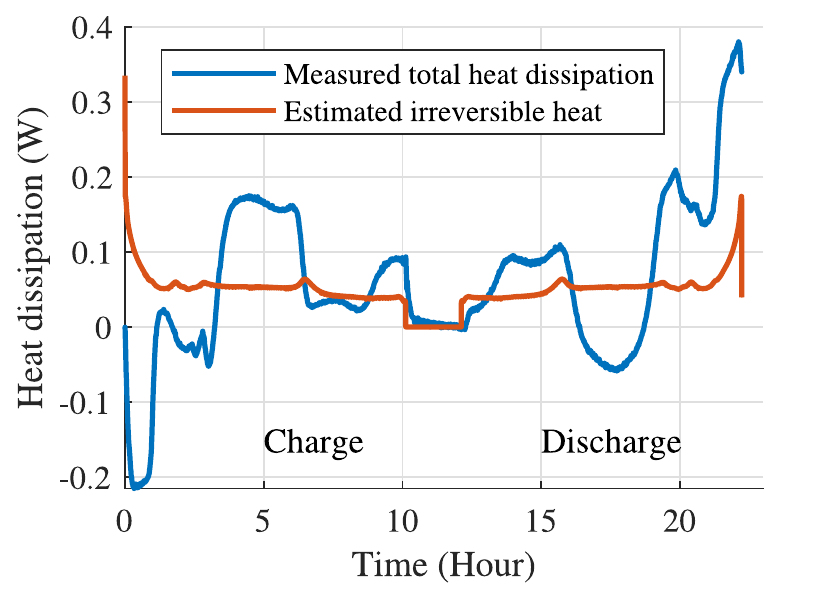}
         \caption{}
          \label{fig:Method2_measurement}
     \end{subfigure}
     \hfill
        \caption{(a) and (b) The experiment setup of the calorimeter method. (c) The measured charge and discharge voltage curve and the estimated open circuit potential at 20 $^{\circ}$C. (d) The total heat dissipation during a 0.1 C charge and discharge cycle is measured in the calorimetric method at 20 $^{\circ}$C.} 
        \label{fig:Method2_setup}
\end{figure*}

The result from the potentiometric method is shown in Fig.~\ref{fig:Method1_result} and there is no notable difference between the two battery cells. The entropic coefficient of the measured battery cell is within $\pm$ 0.3 mV~K$^{-1}$ so the voltage change caused by the temperature is much smaller compared with the battery voltage range (2.8~V-4.15~V). Therefore a high accuracy voltage measurement is one of the key factors for obtaining a good result. Moreover the noise level in this measurement is very low in both the temperature and voltage measurement, which gives a good signal to noise ratio.

The potentiometric method is time consuming, both due to the relaxation time and the thermal equilibrium time. Even with the relaxation background subtraction, the battery cell needs to relax for a considerable time before conducting the measurement. For a smaller battery cell this method is more applicable. However, with an increasing battery size, the experiment time increases dramatically which makes this method less favorable.

\subsection{Result when using the Calorimetric method}

Compared with the time consuming potentiometric method, the calorimetric method is more suitable to characterize a large-scale battery cell. The advantage of the calorimetric method is not only fast but also it gives a continuously entropic coefficient profile.

An in-house designed isothermal heat conduction calorimeter is used in this study, as shown in Fig.~\ref{fig:calorimeter1} and \ref{fig:calorimeter2}. This calorimeter is designed for prismatic and pouch type battery cells from 10 cm $\times$ 10 cm and upwards. In the calorimetric method experiment, the entropic coefficient is measured at different current levels and different temperatures (10 $^{\circ}$C, 20 $^{\circ}$C and 30 $^{\circ}$C). To reduce the heat leakage from the cable, the current needs to be regulated to be as low as possible but still be able to produce a significant measurable heat to offer a good signal to noise ratio. Therefore 0.1~C and 0.05~C are chosen to be used in this study.

The calorimeter can only measure the total heat generation which needs to be separated into irreversible heat and reversible heat. In previous studies, the battery impedance measured from electrochemical impedance spectroscopy or current interruption measurement has been used as an assistance to estimate the overpotential \cite{hong1998electrochemical,al2000characterization,nieto2013thermal,vertiz2014thermal,schuster2015thermal,xiao2013theoretical,chen2014measurements}. In this study, the irreversible heat is calculated directly from the battery overpotential which is obtained from the difference between the measured terminal voltage and the open circuit potential estimated from coulomb counting, as shown in Fig.~\ref{fig:Method2_ocv}. 

\begin{figure}[!ht]
     \centering
     \begin{subfigure}[b]{0.48\textwidth}
         \centering
         \includegraphics[width=\textwidth]{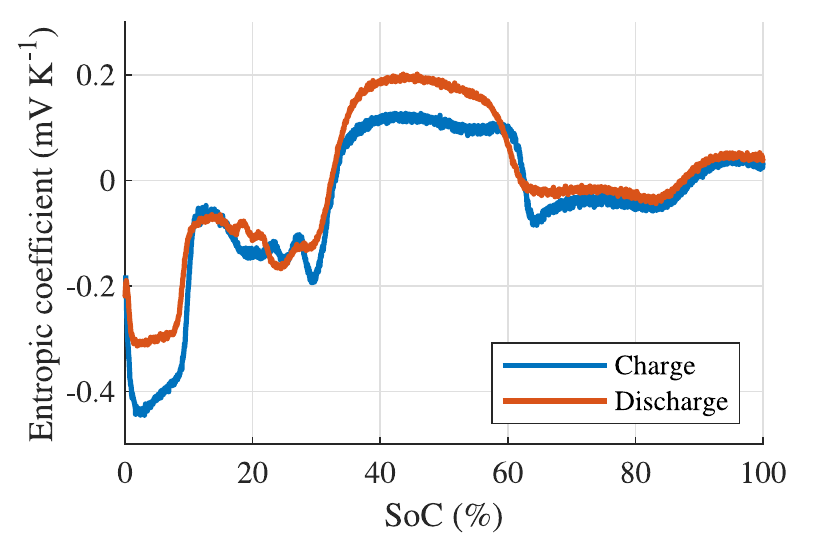}
         \caption{}
         \label{fig:Method2_result_C20_20degree}
     \end{subfigure}
     \hfill
     \begin{subfigure}[b]{0.48\textwidth}
         \centering
         \includegraphics[width=\textwidth]{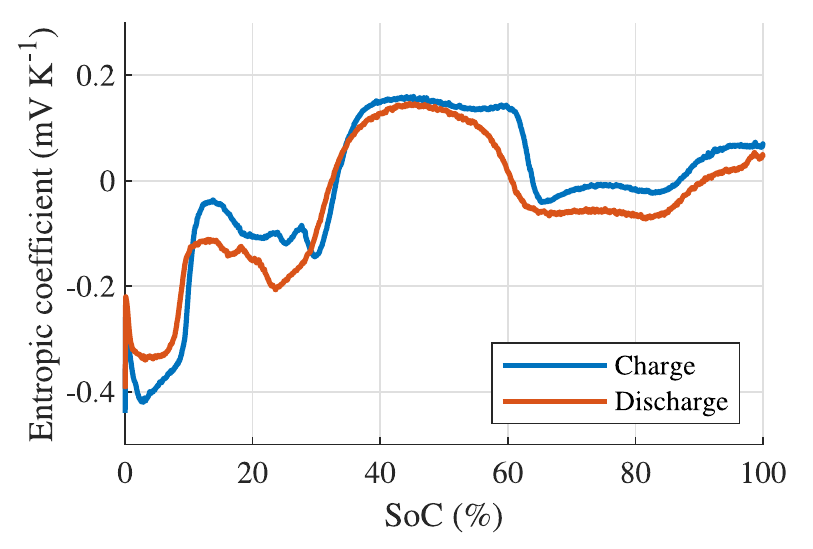}
         \caption{}
         \label{fig:Method2_result_C10_20degree}
     \end{subfigure}
     \hfill
        \caption{Entropic coefficient identified by the calorimetric method at 20 $^{\circ}$C with current level of (a) 0.05~C and (b) 0.1~C.}
        \label{fig:Method2_result}
\end{figure}

In the experiment, the battery cell starts from 0\% SoC and then is charged to 100\% SoC. After 2 hours relaxation, the battery is discharged back to 0\% SoC. One example of the measurement result at 20 $^{\circ}$C with 0.1C current level is shown in Fig.~\ref{fig:Method2_measurement}. As shown, at 0.1C the reversible heat is the main heating source. The total heat generation is reversed between the charge and discharge with a small offset which is the irreversible heat. The irreversible heat is estimated from the overpotential which is the difference between the measured voltage and the open circuit potential.

With the calorimetric method, the measured entropic coefficient profiles at 20 $^{\circ}$C, 0.05~C and 0.1~C are presented in Fig.~\ref{fig:Method2_result} showing a good agreement. The measured entropic coefficient is less affected by the temperature but shows a slight difference at different current levels. Moreover there is a shift observed between the entropic coefficient under charging and discharging. One reason is that the measured heat (Fig.~\ref{fig:Method2_measurement}) has a small delay from the actual heat generation due to the thermal time constant of the battery and the calorimeter. When the discharge curve is flipped over, the delay is in the opposite direction which causes the shift in Fig.~\ref{fig:Method2_result}. Overall the calorimetric method is fast and accurate with repeatable and reliable results. It gives a continuous entropic coefficient profile which is beneficial for the thermal modelling since interpolation is not needed and more details are captured. However, this is obtained with a high cost of the special equipment which is required for the calorimetric method.


\section{Comparison and discussions}

\begin{figure*}[!ht]
    \centering
    \includegraphics[width=0.95\textwidth]{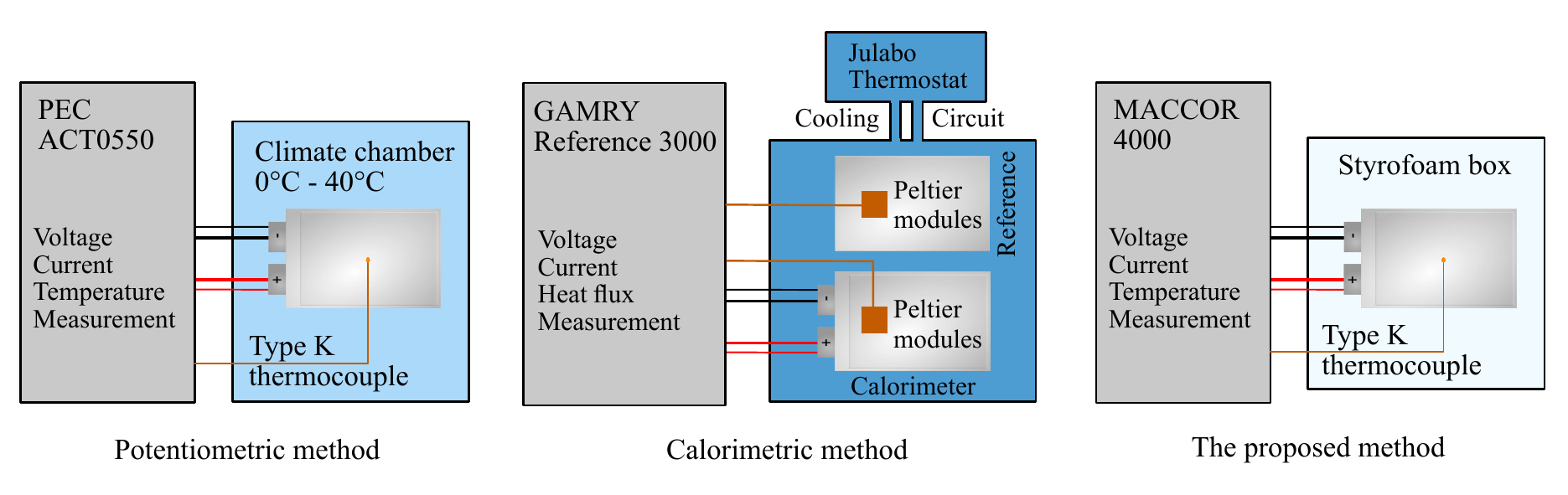} \\
    \caption{Schematics of three methods to measure the entropic coefficient.}
    \label{fig:Methods_illustratoin}
\end{figure*}

\begin{table*}[!ht]
\begin{center}
\footnotesize
\caption{A comparison of three methods to measure the battery entropic coefficient.}
\begin{tabular}{| p{3cm}| p{3.8cm}| p{3.8cm}| p{3.8cm} | }  \hline

 & The potentiometric method
 & The calorimetric method
 & The proposed method \\ \hline

Equipment 
& PEC ACT0550  \newline Programmable climate chamber Espec LU-124 \newline One type K thermocouple
& GAMRY Reference 3000 \newline An in-house designed calorimeter with thermostat from Julabo and 6 Peltier modules GM250-127-14-10
& MACCOR 4000 \newline One type K thermocouple \\ \hline

Accuracy and resolution of measurements
& Voltage: $\pm$0.005\% Full Scale Deviation accuracy \newline Voltage: 16 bits resolution \newline Temperature 0.01 $^o$C resolution
& Thermostat temperature instability: 0.01 K  
& Voltage: $\pm$0.02\% Full Scale Deviation accuracy \newline Voltage: 16 bits resolution \newline Temperature 0.01 $^o$C resolution  \\ \hline

Sampling frequency
& 0.2 Hz
& 0.2 Hz 
& 1 Hz  \\ \hline

Performance versus time
& 190 hours for 11 SoC points 
& 20 hours for a continuous curve 
& 16 hours for 9 SoC intervals  \\ \hline

Cost
& Medium
& High 
& Low  \\ \hline

Suitable cell format
& Small cell in any format 
& Pouch cell or prismatic cell 
& Large cell in any format  \\ \hline

\end{tabular}
\label{tab:compare}
\end{center}
\end{table*}

\begin{figure}[!ht]
    \centering
    \includegraphics{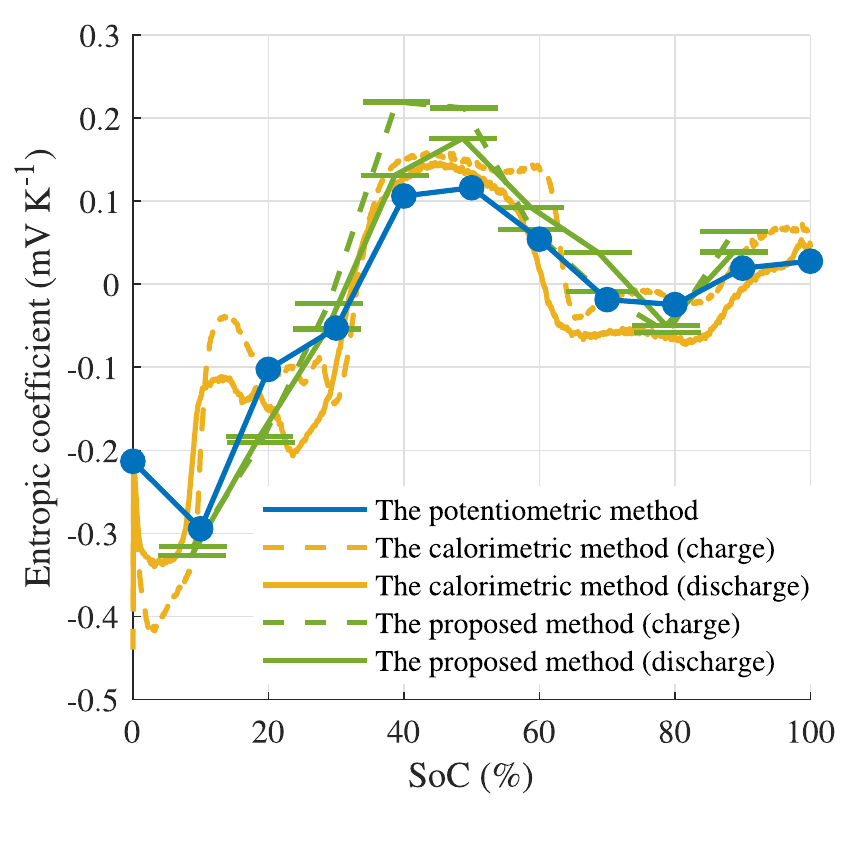} \\
    \caption{Summary of the entropic coefficient measured by the three methods: the potentiometric method, the calorimetric method and the proposed alternating current method.}
    \label{fig:Method_all_result}
\end{figure}

The entropic coefficient measured with the potentiometric method, the calorimetric method and the proposed alternating current method are summarized in Fig.\ref{fig:Method_all_result}, with a good agreement.

In the proposed method, it has been assumed that the battery impedance is constant for the SoC interval with both directions of the current. With this assumption, it is possible the conclude that the observed temperature swing is only due to the entropy effect. However in reality the battery impedance is normally higher at low state of charge and this difference could introduce a different power dissipation and thus a resulting variation in temperature swing. To avoid the impact of various impedance within the SoC interval, it is beneficial to use a small SoC interval in the proposed method. Moreover, the battery impedance can be affected by the direction of the current which has not been sufficiently studied. Nonetheless, the result in Fig. \ref{fig:Method_all_result} still shows a good agreement indicating that the assumption is valid. The thermal model used in the proposed method is a simple lumped parameter network as shown in (\ref{eq:Heat_transfer}). This simplification is made since the system behaviour could be modelled with a first order system during the calibration test. A more complicated thermal model could be applied, however, the in this work a base thermal model provided a satisfying result. In the proposed method, a SoC interval of 10 \% is used with two current levels and cycling periods (1 C with 12 minutes and 1.5 C with 8 minutes). The cycling period is chosen so that the temperature can develop to a distinguishable level, especially in the system with a large thermal capacity. In general it is desired to determine the entropic coefficient with a finer SoC step, which can be achieved by a smaller thermal capacitance (quick temperature response) and using temperature sensors with a higher resolution, as well as increasing the number of the cycles to reduce the impact of the background white noise in the frequency domain. When it comes to the application of the proposed method, the caused aging impact is worth to be taken into consideration. A thorough aging study on the same type of battery cell \cite{wikner2017lithium} shows that cycling within a small state of charge interval, especially at lower state of charge, does not accelerate the degradation of the cell. The lifetime of the used cell is a few thousands of full cycles equivalents and in the proposed method the test procedure is only corresponding to 9 full cycle equivalents (10 cycles at each 10 \% SoC interval for 9 intervals) which can be considered to be non-significant.

The experiment setups used in this work are summarized in Fig.~\ref{fig:Methods_illustratoin} and Table~\ref{tab:compare} for the three methods. The experiments were performed in different institutions and therefore the available test facilities utilized differed. However the test results are not limited due to the specific hardware used. Each method has its advantages and disadvantages and are suitable for different applications. The potentiometric method is straight forward and easy to implement but it requires an environment with controllable temperature and high precision voltage measurement. Moreover it is very time consuming and can only measure the entropic coefficient at individual SoC levels. The calorimetric method is much more time efficient and it gives a continuous profile of the entropic coefficient which is valuable. The main limitation of the calorimetric method is the high cost of the equipment as well as the fact that a battery single cell needs to be available to perform the measurement. The proposed alternating current method targets the large-scale battery cells with the most flexible experimental setup and provides a short experiment time. In principle, a standard vehicle battery with the built in temperature sensors can be used directly without modifications. A drawback is that only the average entropic coefficient within a SoC interval can be measured.

\newpage

\section{Conclusion}
In this paper, a novel method was demonstrated to determine the entropic coefficient for a large commercial lithium ion pouch cell. In this method an alternating current was applied within a small SoC range to create a temperature swing, and thus the entropic coefficient can be obtained by Fourier analysis. This method is fast and require a low effort to set up the experiment. The result obtained from the proposed method was also compared with the potentiometric and calorimetric methods with a very good agreement. 

The current used to determine the entropic coefficient ranged from zero up to 1.5 C, and they gave similar results. It shows that the C-rate has little impact on the measured entropic coefficient, even though there is a slight shift between charging and discharging observed in the calorimetric method and the proposed alternating current method.

The traditional potentiometric method is time consuming and it requires high precision voltage measurement devices. Moreover, it can only measure the entropic coefficient at a single SoC level at a time for each test. The calorimetric method on the other hand could give a continuous profile of the entropic coefficient in the whole SoC range which however, requires an even more dedicated setup to perform the experiment. The proposed novel method is a fast method which measures the average entropic coefficient within a small SoC range. Furthermore, it has a potential to be implemented in a real vehicle since it does not require any extra device other than the normal on-board battery management system.


%



\section*{Acknowledgment}

The authors would like to thank Energimyndigheten, AB Volvo and Volvo Car Corporation for the financing of this work.

\ifCLASSOPTIONcaptionsoff
  \newpage
\fi



%

\bibliographystyle{IEEEtran}
\bibliography{reference.bib}

%

\begin{IEEEbiography}[{\includegraphics[width=1in,height=1.25in,clip,keepaspectratio]{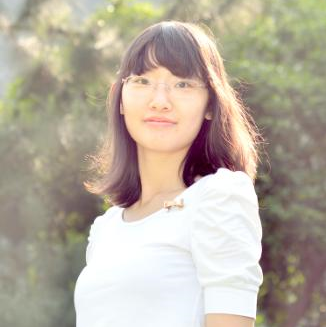}}]{Zeyang Geng}

Zeyang Geng is a PhD student at Chalmers University of Technology working with characterization, modelling and aging of lithium ion batteries. She took her B.Sc at Zhejiang University in 2013 and M.Sc at Chalmers University of Technology in 2015.
\end{IEEEbiography}

\begin{IEEEbiography}[{\includegraphics[width=1in,height=1.25in,clip,keepaspectratio]{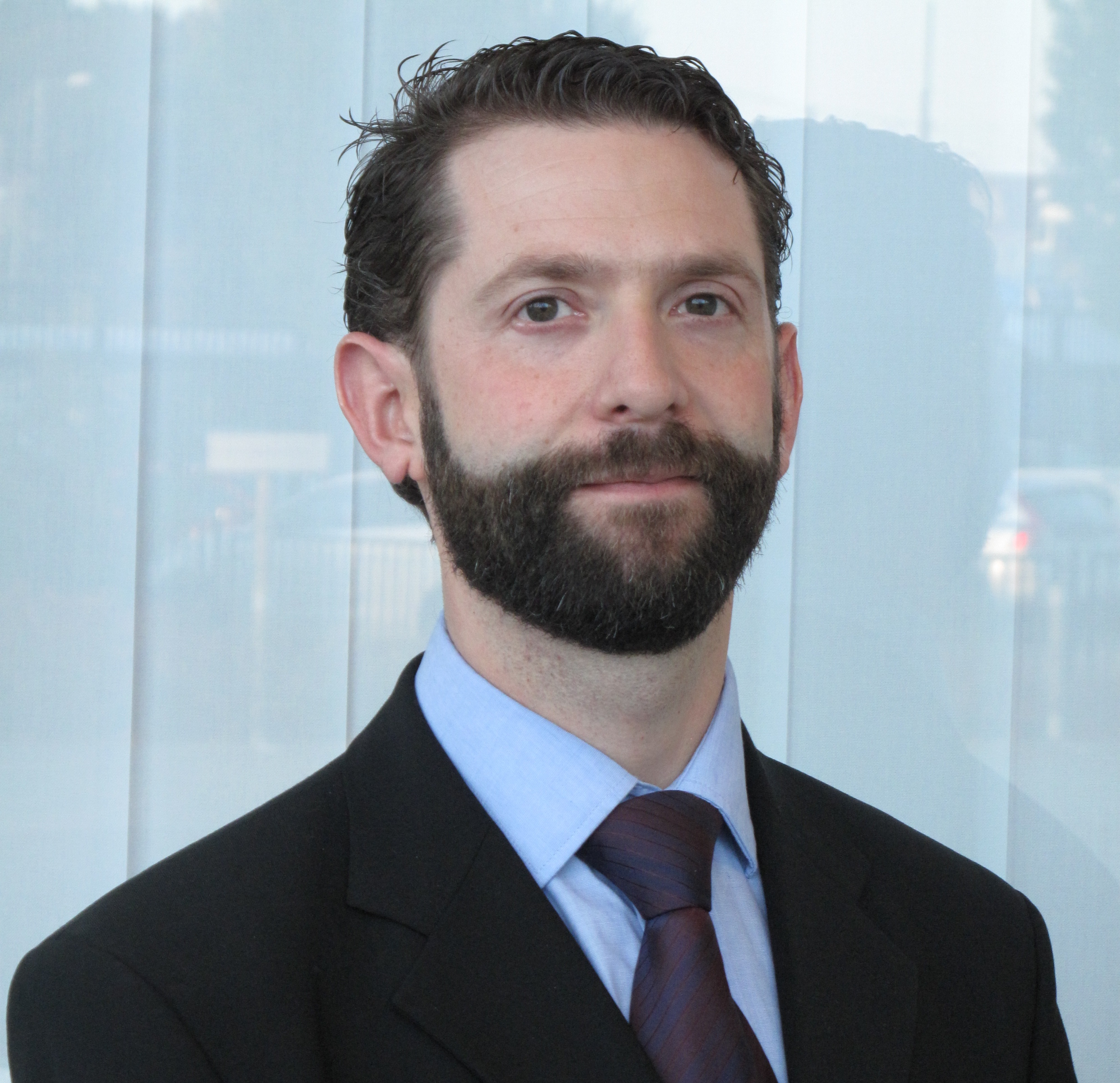}}]{Jens Groot}

Dr. Jens Groot has worked within research and development for heavy-duty vehicles with the Volvo Group since 2003 focusing on battery characterization, evaluation and modelling. Since 2012 he holds a position as energy storage specialist primarily working with cycle life evaluation and cell modelling in both thermal and electrochemical domain. He received his Ph.D. from Chalmers University of Technology, where the main focus of his research was on high-power Li-ion battery lifetime prediction and modelling.
\end{IEEEbiography}

\begin{IEEEbiography}[{\includegraphics[width=1in,height=1.25in,clip,keepaspectratio]{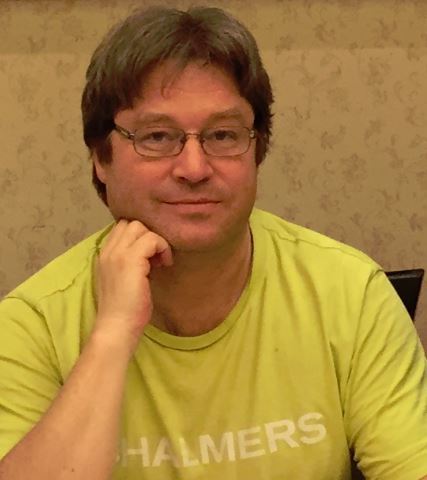}}]{Torbjörn Thiringer}

Torbjörn Thiringer works at Chalmers University of Technology, in Göteborg Sweden, as a professor in applied power electronics. He took his M.Sc and Ph.D at Chalmers University of Technology in 1989 and 1996 respectively. His areas of interest include the modeling, control and grid integration of wind energy converters into power grids, battery technology from cell modelling to system aspects, as well as power electronics and drives for other types of applications, such as electrified vehicles, buildings and industrial applications.
\end{IEEEbiography}







\end{document}